\documentclass[12pt,preprint]{aastex}

\def\gs{\mathrel{\raise0.35ex\hbox{$\scriptstyle >$}\kern-0.6em
\lower0.40ex\hbox{{$\scriptstyle \sim$}}}}
\def\ls{\mathrel{\raise0.35ex\hbox{$\scriptstyle <$}\kern-0.6em
\lower0.40ex\hbox{{$\scriptstyle \sim$}}}}

\shortauthors{Owen et al}
\shorttitle{Deep SWIRE Field III.}

\begin{document}

\title{The Deep SWIRE Field}
\title{III. WIYN Spectroscopy}

\author{
Frazer\,N.\ Owen,\altaffilmark{1,2}, 
\& G.\,E.\ Morrison,\altaffilmark{2,3,4}}
\altaffiltext{1}{National Radio Astronomy Observatory, P.\ O.\ Box O,
Socorro, NM 87801 USA.; fowen@nrao.edu The National Radio Astronomy
Observatory is facility of the National Science Foundation operated
under cooperative agreement by Associated Universities Inc.}
\altaffiltext{2}{Visiting astronomer, Kitt Peak National Observatory,
National Optical Astronomy Observatories, operated by AURA, Inc.,
under cooperative agreement with the National Science Foundation.}
\altaffiltext{3}{Institute for Astronomy, University of Hawaii,
  Honolulu, 96822, USA}
\altaffiltext{4}{Canada-France-Hawaii Telescope, Kamuela, Hawaii, 96743,USA}

\setcounter{footnote}{5}

\begin{abstract}

	We present the results of spectroscopy using HYDRA on the 
WIYN 3.5m telescope of
objects in the deep SWIRE radio field. The goal of the project was
to determine spectroscopic redshifts for as many of the brighter
objects in the field as possible, especially those detected in the
radio and at 24$\mu$m. These redshifts are primarily being used in
studies of galaxy evolution and the connection of that evolution to
AGN and star-formation. 
Redshifts measured for 365 individual
objects are reported. The redshifts range from 0.03 to 2.5, mostly
with $z < 0.9$.   The sources were selected to
be within the WIYN HYDRA field of approximately 30\arcmin\ in radius from the
center of the SWIRE deep field,  10$^h$46$^m$00$^s$, 
59\arcdeg01\arcmin00\arcsec\ (J2000). 
Optical sources for spectroscopic observation were selected from a
r-band image of
the field.  A priority list of spectroscopic targets was established
in the following order: 20cm detections, 24µm detections, 
galaxies with $r < 20$ and the balance made up of fainter galaxies
in the field.
We provide a table listing the galaxy
positions, measured redshift and error, and note any emission lines
that were visible in the spectrum.  In practice almost all the
galaxies with $r < 19$ were observed including all of the radio
sources and most of the 24$\mu$m sources with $r < 20$ and a sample of
radio sources which had fainter optical counterparts on the r-band
image. 

\end{abstract}

\keywords{cosmology: observations ---  galaxies:
evolution --- galaxies: starburst --- infrared: galaxies
galaxies}

\section{Introduction}

This paper is the third in a series documenting our study of the
deep SWIRE field centered at 10$^h$46$^m$00$^s$, 
59\arcdeg01\arcmin00\arcsec (J2000). Paper I \citep{o20} describes
the 20cm VLA observations which produced the deepest 20cm radio survey
to date with 2050 sources and the basic radio properties of the faint $\mu$Jy
population. Paper II \citep{o90} details a complementary, deep
90cm survey and dependence of 20cm to 90cm spectral index
on radio flux density. These two papers concentrate only on
radio properties. What is needed to learn more are the
distances to the sources and their properties at other wavelengths.
In paper IV \citep{p09} we discuss the absolute radio/optical/NIR properties of
the radio sources and their connection to the evolution of galaxies.
We accomplish this goal by estimating photometric redshifts for the
sources for which we do not possess a spectroscopic redshift, using
images of the field from the GALEX uv bands, ground-based UgrizJHK and
the 3.6 and 4.5 $\mu$m bands of IRAC on Spitzer. As part of our
photometric technique we need spectroscopic redshifts as a training
set. The results described in this paper are used in paper IV, along
with redshifts from other projects for this purpose. These
spectroscopic redshifts also
provide secure, accurate redshifts, and thus distances, for a large
fraction of the low redshift objects in the field. 

The results reported here have already been used as calibration
data for another photometric redshift study \citep{rr08}, 
which has itself been used
for in a number of subsequent studies. A secondary goal of the
survey was to observe 24$\mu$m sources and those redshifts were
used for a determination of the 24$\mu$m luminosity function
\citep{o06}. Moreover these spectroscopic and the photometric
redshifts from paper IV are 
being used for a number of studies of the AGN versus
star-formation activity versus redshift. In addition, deep Spitzer MIPS imaging
is being stacked as a function of radio luminosity and redshift
in order to use the radio-FIR relation to indicate the relative
evolution of these two properties. The redshifts are also being used
with the Chandra survey of this region to study the evolution of
the X-ray-loud AGN population and its connection to the radio
population \citep{bw}. Several other studies that are currently underway are
also making use of the WIYN redshifts and the photometric redshifts 
in order to fold the deep radio data into a more complete picture of
galaxy evolution provided at other wavelengths.

\section{Selection, Observations and Reduction}

The sources were selected to be within the WIYN HYDRA field of
view, approximately 30\arcmin\ in radius from  the center of the SWIRE
deep field, 10$^h$46$^m$00$^s$, 59\arcdeg01\arcmin00\arcsec\ (J2000).
Optical sources for observation were selected the r-band image
of the field taken for the SWIRE survey
with the KPNO 4m \citep{l03}. The magnitudes were measured using a
2\arcsec\ diameter aperture, which is the
same size as the fibers. A preliminary 20cm radio
image from the VLA and a preliminary 24$\mu$m from Spitzer existed
at the time of the observation. A priority list of possible sources
was established:  20cm detections , 24$\mu$m
detections, galaxies with $r<20$, and fainter galaxies ($r>20$) in the field.
As described below sources for observations were optimized for each
fiber set.

The observations were made Feb 11-15, 2004 using the HYDRA
fiber spectrograph on the WIYN 3.5m telescope at KPNO. 
The $316@7.0$ grating was used with the Bench Spectrograph and
the red fiber cables. The G4 $GG-420$ blocking filter was used
which after calibration yielded a usable spectral region of
4400\AA\ to 9700\AA\ with some second order contamination beyond
8800\AA. 
Two CuAr comparison lamp exposures were taken for each wavelength
calibration, one with $10 \times$ the integration
time of the other, were observed before and after each
fiber set. These data showed that the wavelength response of
the bench spectrograph setup was very stable. Standard stars were 
observed each night though single fibers to allow a rough relative spectral
calibration to be obtained. Radial velocity standards were
also observed to check the stability and accuracy of the redshift
measurements. These observation demonstrated that the overall
calibration of our unknown sources were limited by other
considerations than the calibration, mainly S/N. 

The fiber setup was done using {\it whydra}. A large number of
runs of the program with slightly different parameters were
used to obtain the best fiber setup for each observation. Typically, 
20 sky fibers, 5 FOPS (guide stars) and 65 objects fibers were used.
Depending on observing conditions different criteria were used
to prioritize the setup, {\it e.g.} brighter objects were observed
preferentially when the transparency and/or seeing was poorer. Total 
observing times
of one to four hours were used, usually split into four shorter
exposures in order to allow easier removal of cosmic rays. 
We observed 12 separate, useful fiber configurations with nearly the
same pointing center but slightly shifted to reduce the number
of the fiber collisions which limited the number of sources we could
observe at once. The weather and other conditions, clouds, seeing,
position of the moon, varied as usual over
the run and we worked our way through our priority list. We were
able to examine the spectra in real time and evaluate how well we
were doing. We used this information as we went along to plan the
next fiber configuration. For example if the transparency and/or
seeing were likely to be poor during a fiber set we picked brighter,
sometimes low priority objects, to observe. At times we observed a
brighter set of objects with a shorter total observing time in order
to have all the observations of a given fiber set fit into the time
during a single night. In the end, these constraints made the data
quality of the spectra somewhat variable and at times we lost high
priority objects entirely.

The IRAF {\it ccdproc} package was used for the reductions. First
the observations were bias corrected and then the {\it dohydra}
script was used. Dome flats were used both to determine 
the locations of the fibers on the CCD and for flat fielding. The
two comparison lamp exposures were combine so that good S/N could
be obtained in both the red and blue parts of the spectrum. Bad
sky frames were edited out during the procedure. 
The individual spectra were extracted to a one-dimensional format and then
combined using the IRAF task, {\it scombine}. The final spectra
were then edited using {\it splot}. Redshifts were derived either
by calculating equivalent widths and center wavelengths for
individual lines, mostly emission lines, in each spectrum
using {\it splot}
or cross-correlating absorption-line spectra with standard spectra
using the IRAF task {\it fxcor}.
Either a radial velocity standard or elliptical galaxies obtained
in previous observing runs with wider spectral coverage and better
spectral resolution were used as the template spectrum. Errors were
estimated by combining the internal errors from the cross-correlations
for absorption spectra, or the scatter in the redshift estimates for
multiple lines in quadrature with 50 km s$^{-1}$, the upper limit to
the error in the absolute spectral calibration derived from the
scatter in velocity derived for observations of radial velocity
standard stars. 

\begin{figure}
\epsscale{0.9}
\plotone{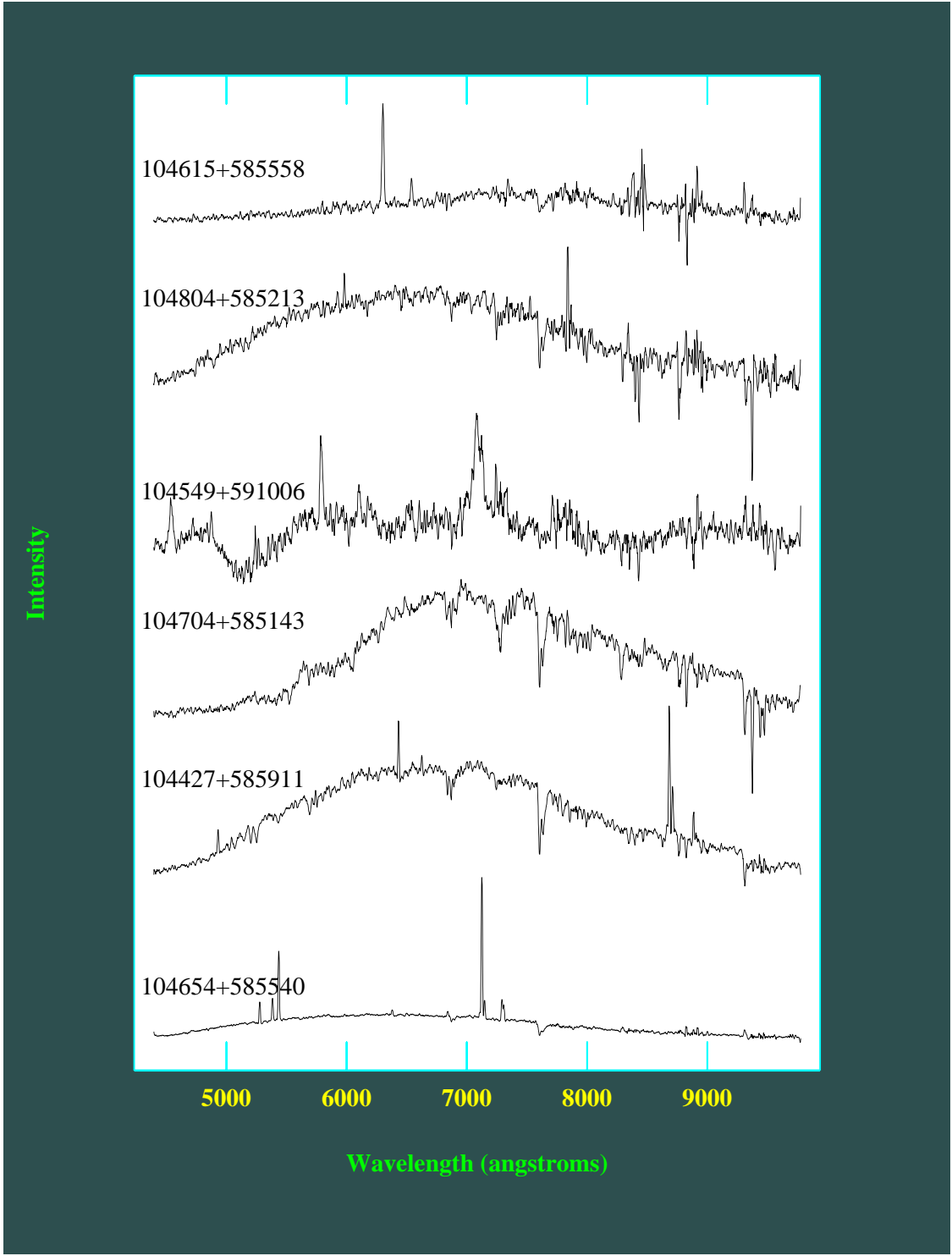}
\caption{Examples of the observed, uncalibrated spectra used to
  estimate redshifts. $104615+585558$: $z=0.68$, narrow emission lines;
$104804+585213$: $z=0.19$, weak emission lines; $104549+591006$:
  $z=2.75$, broad emission lines; $104704+585143$: $z=0.41$, old stellar
population absorption lines; $104427+585911$: $z=0.32$, narrow
  emission lines; $104654+585540$: $z=0.12$, strong emission
  lines. See table~\ref{reds} for details of lines used for redshifts.
 \label{spec}} 
\end{figure}

In figure~\ref{spec}, we show six examples of spectra we used in
to determine the results in tables. These spectra are
picked to cover the range of different types of objects in our sample
from broad-line, high redshift AGN to pure absorption-line
spectra. One can see the increasing importance of night sky features
at longer wavelengths but that strong emission lines can still be seen in
the red. 
After completing the reductions out of 1063 objects in our original sample, 548
had a least one reduced spectrum. Of these we obtain reliable
redshifts for 365 objects. For the remaining 183 we could not measure 
a robust 
redshift. In most of these cases the signal in the spectrum was too
weak to be usable due to poor observing conditions. In
table~\ref{count} we summarize the statistics of the observed sources
as a function of r magnitude.

\section{Results}

In table~\ref{pos}, we give the first ten columns of
the electronic table containing J2000 positions and r magnitudes of
the objects with for which we obtained successful spectra from WIYN. 
In table~\ref{reds}, we present the first ten columns of the
electronic table containing the results of the WIYN spectroscopy. 
In columns
1 and 2, we give the J2000 coordinate name and the radio source
name. If column 2 is blank that means that the galaxy was not detected
in the five sigma 20cm catalog given in paper I. 
Columns 3 and 4 contain the redshift and error. In column 5 we give
the ``notes'' on the results, including either the spectral line(s) and/or the
``FXCOR'' if the redshift was obtained by cross-correlation with an
absorption-line template. When a photometric redshift is used to confirm a one
line redshift we add the note ``photoz'. If the source was detected at
20cm at the three sigma or four sigma level we also note that in column
5. Finally, we give the number of the galaxy in the input file for the
WIYN observation for future reference and requests for individual
spectra.

\begin{figure}
\plotone{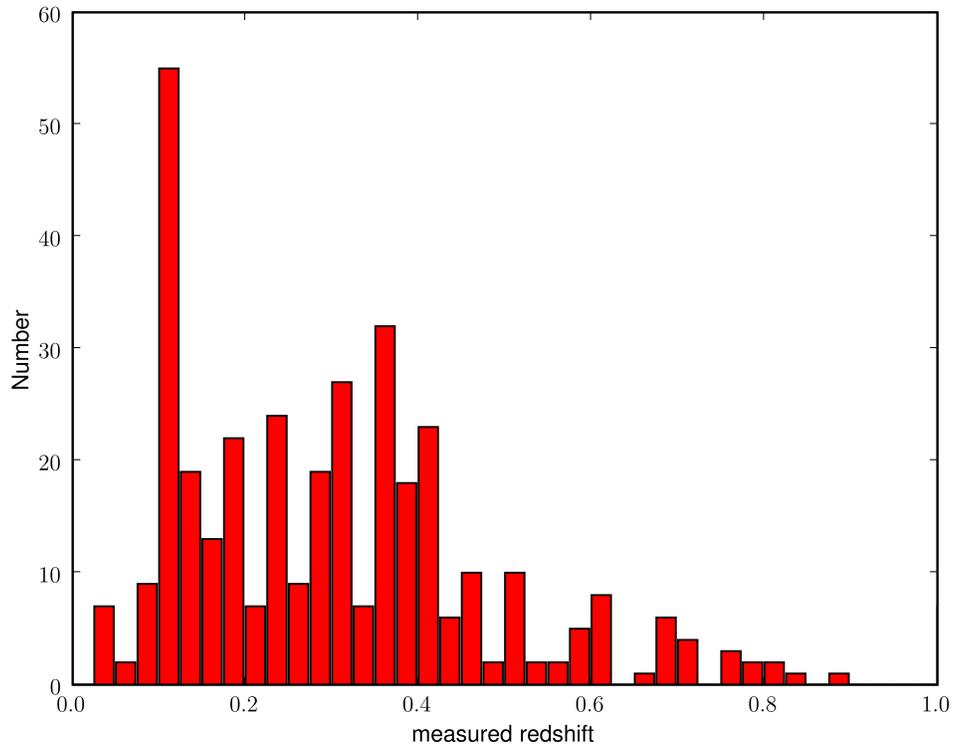}
\caption{Histogram of redshift distribution $0.0<z<1.0$. Eight objects
have measured redshifts in the range $1.0<z<2.75$. \label{hist}} 
\end{figure}

Of the 365 total objects for which we measured a redshift, 
267 of the objects have emission lines; 249 have two or more
lines. For the 18 objects with only one line, the identity of the emission
line (usually OII) is confirmed by an absorption line redshift in
nine cases and
in the other nine cases by a photometric redshift from \citet{p09}. 98 objects
have measured redshifts from absorption lines only. 
249 of the measured redshifts are for radio detected objects. 229
are published in the five sigma catalog in paper I. Ten are detected
at the four sigma level and ten more at three sigma.  286 24m$\mu$
sources had their redshifts determined \citep{o06}.  
In Figure~\ref{hist} we show the redshift distribution. The largest
peak contains 53 objects with $0.11<z \le 0.12$ suggesting we are
looking through an organized structure at this redshift, although no
concentrated cluster is obvious in the images. Eight objects not
shown have redshifts $>1$.

\clearpage

\begin{deluxetable}{rrrr}
\tablecolumns{4}
\tablewidth{0pt}
\tablecaption{Observed Source Statistics \label{count}}
\tablenum{1}
\pagestyle{empty}
\tablehead{
\colhead{r Mag}& 
\colhead{\# in cat}&
\colhead{\# observ}&
\colhead{\# z meas}}
\startdata
$r\le 19$&73&72&67\\
$19<r\le 20$&218&176&129\\
$20<r\le 21$&633&189&125\\
$21<r\le 24$&139&111&44\\
all&1063&548&365\\

\enddata
\end{deluxetable}
\clearpage

\begin{deluxetable}{rrrr}
\tablecolumns{4}
\tablewidth{0pt}
\tablecaption{Observed Source Positions\label{pos}}
\tablenum{2}
\pagestyle{empty}
\tablehead{
\colhead{Name}& 
\colhead{RA(2000.0)}&
\colhead{Dec(2000.0)}&
\colhead{r Mag}}
\startdata
104340+591241&10 43 40.32&59 12 41.1&19.3\\
104340+591826&10 43 40.77&59 18 26.4&19.0\\
104341+591733&10 43 41.38&59 17 33.7&19.9\\
104341+591654&10 43 41.50&59 16 54.0&18.3\\
104342+591158&10 43 42.22&59 11 58.2&19.4\\
104343+585535&10 43 43.21&58 55 35.7&19.9\\
104347+584756&10 43 47.17&58 47 56.9&20.1\\
104347+590217&10 43 47.61&59 02 17.3&19.5\\
104347+590620&10 43 47.81&59 06 20.9&19.3\\
104351+585544&10 43 51.24&58 55 44.2&21.0\\

\enddata
\end{deluxetable}
\clearpage

\begin{deluxetable}{rrrrl}
\tablecolumns{5}
\tablewidth{0pt}
\tablecaption{Redshifts and Associated Data\label{reds}}
\tablenum{3}
\pagestyle{empty}
\tablehead{
\colhead{Coord}& 
\colhead{Radio}&
\colhead{z}&
\colhead{error}&
\colhead{Notes}\\
\colhead{Name}&
\colhead{Name}&
\colhead{}&
\colhead{z}&
\colhead{}}
\startdata
104340+591241&       &0.1994&0.0002& OIII, NII, Ha, Hb, 4sigma 7077\\
104340+591826&       &0.0281&0.0002& NII, SII, Ha 7043\\
104341+591733&       &0.2286&0.0002& OIII, NII, Ha, Hb 7186\\
104341+591654&       &0.0286&0.0002&OIII, OI, NII, SII, HeI, Ha, Hb 7016\\
104342+591158&       &0.1994&0.0002&FXCOR 7087\\
104343+585535&       &0.4714&0.0002&OII, OIII, Hb, Hg, Ha, NII Balmer lines broad 7183\\
104347+584756&       &0.4187&0.0003&FXCOR 7250\\
104347+590217&       &0.3223&0.0003&FXCOR 7101\\
104347+590620&10061  &0.1417&0.0002&OIII, OI, NII, SII, HeI, Ha, Hb 4sigma 7076\\
104351+585544&00053  &0.6038&0.0003&OII, OIII, Hb 7741\\

\enddata
\end{deluxetable}
\clearpage

\clearpage

\end{document}